\documentclass[runningheads]{llncs}
\usepackage{graphicx}
\usepackage[symbol]{footmisc}

\begin{document}
\title{DIME: An Online Tool for the Visual Comparison of Cross-Modal Retrieval Models}
%
%
\author{Tony Zhao\inst{1} \and
Jaeyoung Choi\inst{2} \and
Gerald Friedland\inst{1}}
\authorrunning{Zhao et al.}
\titlerunning{DIME}%
%
\institute{University of California at Berkeley, USA \and International Computer Science Institute, Berkeley, USA}
\maketitle              
\begin{abstract}
Cross-modal retrieval relies on accurate models to retrieve relevant results for queries across modalities such as image, text, and video. In this paper, we build upon previous work by tackling the difficulty of evaluating models both quantitatively and qualitatively quickly. We present DIME (Dataset, Index, Model, Embedding), a modality-agnostic tool that handles multimodal datasets, trained models, and data preprocessors to support straightforward model comparison with a web browser graphical user interface. DIME inherently supports building modality-agnostic queryable indexes and extraction of relevant feature embeddings, and thus effectively doubles as an efficient cross-modal tool to explore and search through datasets.

\end{abstract}
\section{Introduction}
	With the exponentially increasing amount of multimedia content on the Internet of different modalities, cross-modal retrieval becomes increasingly important for indexing and searching through either large multi-modal datasets or between datasets of different modalities. Such cross-modal models allow retrieval across modalities such as text, image, video, etc.~\cite{ref1}\cite{ref2}. 
	Standard evaluation metrics such as mean average precision (mAP) allow quantitative comparison between models, however, visual inspection of retrieval results allows qualitative evaluation of the models which provides much practical insight into the behavior of the models.
	Demands for interactive search-and-retrieval tools that can accommodate large-scale datasets have been rapidly increasing in the multimedia community and many such tools~\cite{ref3}\cite{ref4}\cite{deepfeatures} have been released in this regard, or to showcase the underlying cross-modal retrieval models. 
	

	  In this paper, we present DIME (Dataset, Index, Model, Embedding) that builds upon related work in creating systems to index and explore large datasets of images as well as other modalities. For instance, systems such as ImageX~\cite{ref3} share similar functionality such as ranking images by their similarity to a query image or classification concept instead of search through filtering. 
	  Most general-purpose content-based multimedia retrieval stacks such as Vitrivr~\cite{ref4} support two major workflows. Within these workflows, DIME adds several key features that we found to be very useful in cross-modal retrieval research, which are either lacking or missing in previous works. 
      
       First, the \textit{extraction workflow} is used to process different media objects such as images, audio, and video to generate relevant features. DIME adds some minor efficiency improvements and functionality to this workflow such as binarization and bit-packing of embeddings. These improvements allows for memory-efficient extraction of feature vectors and removes the need to repeatedly process datasets while testing models.
      
      Second, the \textit{query workflow} is used when users query a dataset through an interface. This involves extracting the relevant features of the query and dataset look-up before ranking results by their measured similarity. DIME also adds some minor improvements to this workflow by adding metadata and diagnostic information on top of the database look-up. 
      
      Finally, DIME improves the flexibility of these systems by supporting a completely new workflow that we found to be highly useful for cross-modal retrieval research. Since feature-based retrieval relies on accurate models, DIME implements the \textit{model workflow}, which is used to compare models against each other both quantitatively with metrics as well as qualitatively with the visual presentation of results that are returned by queries. 
	
	 On top of these new features, DIME automates the process of building modality-agnostic queryable indexes from corresponding datasets and models as well as presenting information useful for model comparison. Lastly, we implemented a graphical user interface in the web browser to support online and remote functionality.

\begin{figure}
\includegraphics[width=\textwidth]{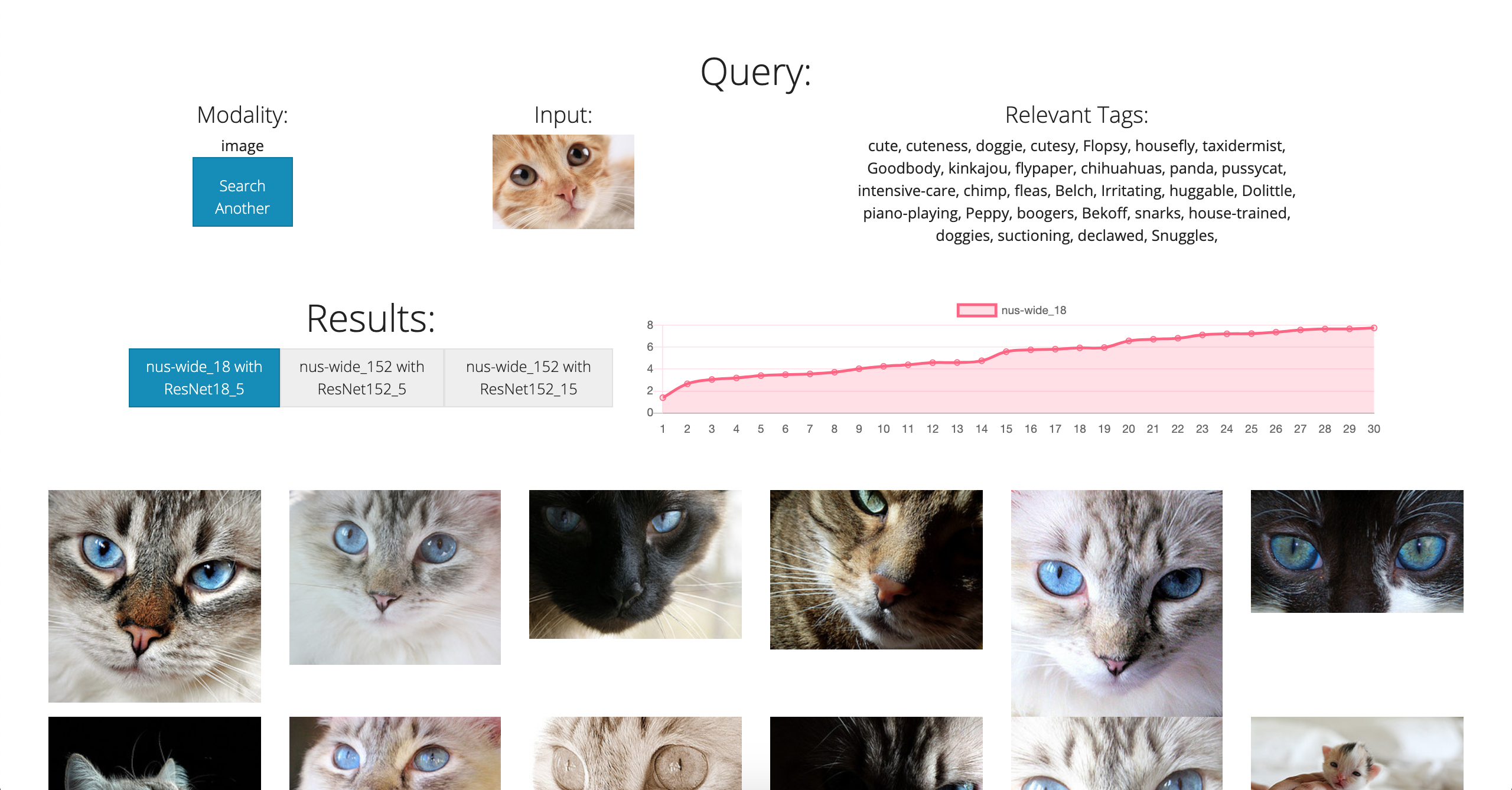}
\caption{DIME provides information between models such as a graph of the distribution of similarity (Euclidean distance) among the results for the query (an image in this figure). The relevant tags are the results returned from a dataset lookup of the text-modality index specific to the relevant model.} \label{fig1}
\end{figure}
    

\begin{figure}
\includegraphics[width=110mm]{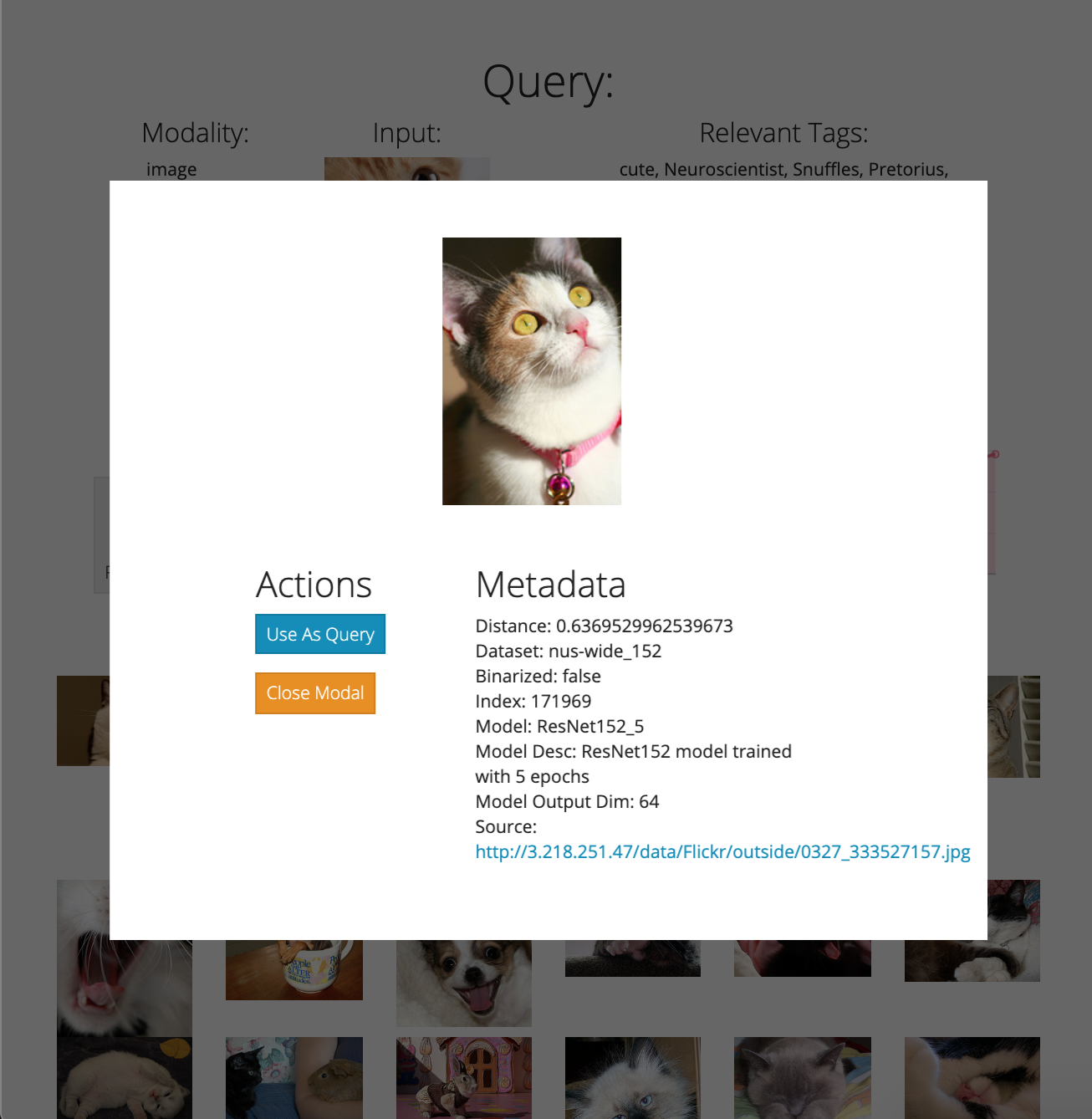}
\caption{DIME provides diagnostic metadata as well as other minor functionality such as using results as queries} \label{fig2}
\end{figure}

\section{DIME: Dataset, Index, Model, Embedding}

DIME is comprised of its titular parts: \textit{dataset}, \textit{index}, \textit{model},  and \textit{embedding}. We define \textit{model} as an object with a callable function. This function has specified input dimensions and output dimensions respectively. A \textit{dataset} is defined as an iterable of vectors. DIME takes a dataset and a model to extract \textit{embeddings} for that dataset from that specific model. Finally, it constructs a queryable \textit{index} from those embeddings to support search and exploration.

DIME receives datasets along with any preprocessors and a specific output dimension. Any pair of model and dataset with matching input and output dimensions respectively, can then be built into an index.  Users can specify if these embeddings should be binarized, which will greatly reduce the amount of memory and time needed to build and load indexes. Once the index has finished building, the entire processed dataset can support queries that pass through modality-appropriate processing.

Database lookup is supported by efficient similarity search. A query such as text or image is first vectorized through appropriate preprocessors to create a query vector. This query vector is then passed through a selected model to extract the query feature embedding. This query feature embedding is then compared against the embeddings in an index using k-nearest neighbors search to calculate similarity (euclidean distance) between the query feature embedding and the embeddings of the dataset. This efficient similarity search algorithm is implemented with FAISS~\cite{FAISS}. The closest $n$ neighbors are returned as the results of the database look-up as well as diagnostic metadata.

\section{Conclusion and Future Work}

To simplify and expedite a tedious process of comparing models for cross-modal retrieval, we present DIME, an online tool that builds upon previous work as a modality-agnostic search engine which automates the construction of queryable indexes from datasets of different modalities.

Besides aesthetic and user interaction improvements, our goal is to make improvements on the quantitative diagnostics of performance and accuracy. A wider array of metrics and tests can be added to help researchers assess and compare models. Likewise, a multitude of diagnostic information and metrics can be implemented for datasets as well. 

Ultimately the end goal of DIME is to remove tedious but currently necessary steps needed to train and compare cross-modal retrieval models. Ideally, a researcher will only need to focus on designing models. In future iterations of the tool, we would like to support automation of training, as the tool currently only supports usage and comparison of pre-trained models.

\section*{Acknowledgment}

		
		Parts of this work was performed under the auspices of the U.S. Department of Energy by Lawrence Livermore National Laboratory under Contract DE-AC52-07NA27344 and was supported by the LLNL-LDRD Program under Project No. 17-SI-003. Computation resources used in this work were partially supported by AWS Cloud Credits for Research. 
		Any findings and conclusions are
		those of the authors, and do not necessarily represent the
		views of the funders.

%
%
%
%

\end{document}